   \documentclass{emulateapj}

\shorttitle{Evidence for a merger of binary white dwarfs: the case of GD 362}
\shortauthors{Garc\'\i a--Berro et al.}

\begin{document}

\title{Evidence for a merger of binary white dwarfs: the case of GD~362}

\author{E. Garc\'\i a--Berro\altaffilmark{1}, 
        P. Lor\'en--Aguilar\altaffilmark{1} and
        A.G. Pedemonte\altaffilmark{1}}
\affil{Departament de F\'\i sica Aplicada, 
       Universitat Polit\`ecnica de Catalunya, 
       Av. del Canal Ol\'\i mpic s/n, 
       E-08860 Castelldefels (Barcelona), 
       Spain}

\email{garcia@fa.upc.edu,
       loren@fa.upc.edu,
       alba@fa.upc.edu}

\author{J. Isern\altaffilmark{1}}
\affil{Institut de Ci\`encies de l'Espai (CSIC), 
       Facultat de Ci\`encies, 
       Campus UAB, Torre C5-parell, 
       E-08193 Bellaterra (Barcelona), 
       Spain}

\email{isern@ieec.fcr.es}

\and 

\author{P. Bergeron,
        P. Dufour and
        P. Brassard}
\affil{D\'epartement de Physique, 
       Universit\'e de Montr\'eal, 
       C.P. 6128, Succ. Centre-Ville, 
       Montr\'eal, 
       Qu\'ebec, 
       Canada H3C 3J7}

\email{bergeron@astro.umontreal.ca, 
       dufourpa@astro.umontreal.ca, 
       brassard@astro.umontreal.ca}

\altaffiltext{1}{Institut d'Estudis Espacials de Catalunya, 
       Ed. Nexus-201, 
       c/ Gran Capit\`a 2--4, 
       E-08034 Barcelona, 
       Spain}

\begin{abstract}
GD~362 is a  massive white dwarf with a  spectrum suggesting a H--rich
atmosphere which  also shows  very high abundances  of Ca, Mg,  Fe and
other  metals.    However,  for  pure   H--atmospheres  the  diffusion
timescales are so short that  very extreme assumptions have to be made
to account  for the  observed abundances of  metals. The  most favored
hypothesis is that the metals are accreted from either a dusty disk or
from an asteroid belt.  Here we propose that the envelope of GD~362 is
dominated  by He,  which  at these  effective  temperatures is  almost
completely  invisible  in  the  spectrum.   This  assumption  strongly
alleviates the problem, since the diffusion timescales are much larger
for He--dominated atmospheres. We  also propose that the He--dominated
atmosphere of  GD~362 is likely  to be the  result of the merger  of a
binary  white  dwarf, a  very  rare event  in  our  Galaxy, since  the
expected galactic rate is $\sim 10^{-2}$~yr$^{-1}$.
\end{abstract}


\section{Introduction}

GD~362 has  been interpreted as  a massive, rather cool  ($T_{\rm eff}
\approx  9740 \pm  50$~K), white  dwarf  with a  heavy accretion  disk
surrounding it  \citep{Kilic,Becklin,Gianinas}. The dusty  disk around
GD~362 produces an excess of infrared radiation which amounts to $\sim
3\%$  of the total  stellar luminosity.   The chemical  composition of
GD~362  is also rather  singular, showing  a hydrogen  rich atmosphere
with  very   high  abundances   of  Ca,  Mg,   Fe  and   other  metals
\citep{Gianinas}.    Thus,  it   is  classified   as  a   massive  DAZ
(hydrogen--rich) white  dwarf.  The  origin of such  particularly high
photospheric   abundances  ---   $\log(N_{\rm   Ca}/N_{\rm  H})=-5.2$,
$\log(N_{\rm   Mg}/N_{\rm   H})=-4.8$   and  $\log(N_{\rm   Fe}/N_{\rm
H})=-4.5$  --- and  of  the  surrounding dusty  disk  around it  still
remains  a mystery.   Since  the diffusion  timescales  for metals  in
H--rich  white dwarfs  are of  only a  few years  \citep{Koester} very
extreme  assumptions  have  to  be  made in  order  to  explain  these
abundances. At present the most widely accepted scenario is disruption
and accretion  of a planetary body,  although for this  scenario to be
feasible  the  planetary system  should  survive  during the  advanced
stages of stellar  evolution, which by no means  is guaranteed.  Thus,
the formation of an asteroid would require the previous existence of a
disk around  this white dwarf  \citep{Livio1,Livio2}.  Particularly, a
recent  analysis \citep{Livio4}  has shown  that planets  around white
dwarfs  with masses  $M_{\rm  WD}  > 0.7  \,  M_{\sun}$ are  generally
expected to be found at orbital  radii $r > 15$~AU because they cannot
survive the planetary nebula phase and that if planets are to be found
at smaller orbital radii around massive white dwarfs, they had to form
as  the  result  of the  merger  of  two  white  dwarfs.  It  is  also
interesting to  note that there  have been previous  suggestions about
white  dwarfs   that  are  merger   products  ---  see   for  instance
\cite{Liebert} --- but these claims  do not have yet any observational
support.

\section{The scenario}

Another possibility is  that some massive white dwarfs  are the result
of  the merger  of a  double white  dwarf close  binary  system.  This
scenario has been studied in several papers. However, in most of these
papers  either   the  resulting  nucleosynthesis   was  not  addressed
\citep{Segretain}, or the spatial resolution was poor \citep{Benz}, or
the   calculations   were   performed   using   crude   approximations
\citep{Mochkovitch}.   Very recently,  and using  a  Smoothed Particle
Hydrodynamics code, a series  of simulations with the adequate spatial
resolution were  performed and the  nucleosynthesis of the  merger was
studied \citep{Guerrero,Loren}.  The  main results of such simulations
are that the less massive white  dwarf of the binary system is totally
disrupted in  a few  orbital periods. A  fraction of the  secondary is
directly  accreted  onto  the  primary  whereas the  remnants  of  the
secondary form a  heavy, rotationally--supported accretion disk around
the primary and little mass  is ejected from the system. The resulting
temperatures are  rather high ($\sim 9\times 10^8$~K)  during the most
violent  phases   of  the  merger,  allowing   for  extensive  nuclear
processing.

\begin{table}[t]
\begin{center}
\caption{Main  results  of the  SPH simulations.}
\begin{tabular}{lcccccc}
\tableline
\tableline
Run & 0.4+0.8 & 0.4+1.0 & 0.4+1.2 & 0.6+0.6 & 0.6+0.8 \\
\tableline
$M_{\rm WD}/M_{\sun}$   & 0.99 & 1.16 & 1.30 & 0.90 & 1.09 \\
$M_{\rm disk}/M_{\sun}$ & 0.21 & 0.24 & 0.30 & 0.30 & 0.29 \\
$M_{\rm ej}/M_{\sun}$   & $10^{-3}$ & $10^{-3}$ & $10^{-3}$ & $10^{-2}$	& $10^{-3}$ \\
\tableline
He  & 0.94 & 0.93 & 0.99 & 0 & 0 \\
C   & $3\times10^{-2}$ & $2\times10^{-2}$ & $5\times10^{-3}$ & 0.4 & 0.4 \\
O   & $1\times10^{-2}$ & $3\times10^{-3}$ & $3\times10^{-3}$ & 0.6 & 0.6 \\
Ca  & $4\times10^{-5}$ & $2\times10^{-4}$ & $9\times10^{-6}$ & 0   & 0   \\
Mg  & $3\times10^{-5}$ & $3\times10^{-5}$ & $6\times10^{-6}$ & 0   & 0   \\
S   & $8\times10^{-5}$ & $2\times10^{-4}$ & $5\times10^{-7}$ & 0   & 0   \\
Si  & $1\times10^{-4}$ & $2\times10^{-4}$ & $3\times10^{-5}$ & 0   & 0   \\
Fe  & $9\times10^{-3}$ & $7\times10^{-3}$ & $5\times10^{-4}$ & 0   & 0   \\
\tableline
\end{tabular}
\end{center}
\end{table}

The  enhancement  of  the  abundances  of the  most  relevant  nuclear
isotopes occurs  when one  of the coalescing  white dwarfs is  made of
pure He.  Table 1 shows  the {\sl average} chemical composition of the
resulting  disk   and  the  main  characteristics   of  some  selected
simulations.  It  should be noted,  however, that the  distribution of
the different elements in  the disk is rather inhomogeneous. Obviously
those parts  of the disk  in which the  material of the  secondary has
been shocked  have undergone  major nuclear processing.   Hence, these
regions are C--  and O--depleted and Si-- and  Fe--enhanced.  In fact,
the innermost regions ($R<0.1 \,R_{\sun}$) of the merged object, which
have approximately the shape of an ellipsoid, are C-- and O--rich.  It
is expected that  this region would be eventually  accreted during the
the first moments of the  cooling phase of the central object, leading
to a  more massive white  dwarf.  We also  find that the  abundance of
intermediate--mass  and iron--group  elements  is considerably  larger
than  that  of  C and  O  in  the  remnants  of the  accretion  stream
\citep{Guerrero} which are at  larger distances, thus favoring smaller
accretion rates in order to explain the Ca abundance.  In any case, if
the photospheric  abundances of GD~362  are to be explained  with this
scenario the accretion of He--rich material is required.

\begin{figure}
\begin{center}
\includegraphics[angle=-90,scale=.30]{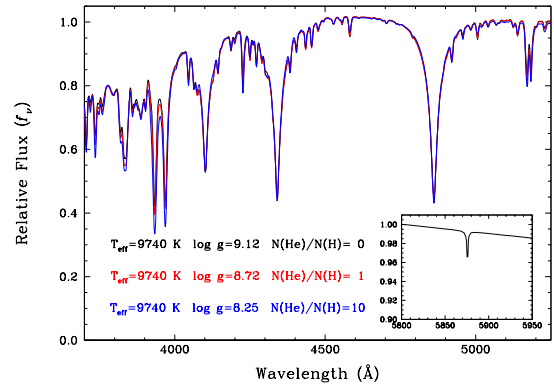}
\caption{Spectrum of GD~362 for three different helium abundances. The
         black line shows the spectrum  of GD~362 when a pure hydrogen
         atmosphere is assumed, leading  to a surface gravity of $\log
         g=9.12$.    For   increasing  amounts   of   He  ---   namely
         $N$(He)/$N$(H)=1,  red  curve,  and  $N$(He)/$N$(H)=10,  blue
         curve  --- the corresponding  surface gravities  are smaller.
         The inset shows an expanded  view of the predicted He line at
         5876~\AA~  for $N$(He)/$N$(H)=10. High  quality spectroscopic
         observations should  be able  to confirm its  presence, which
         has been recently  reported \citep{Jura2}. See the electronic
         edition of the Journal for a color version of this figure.}
\end{center}
\end{figure}

Since He is also accreted onto the surface of GD~362, the photospheric
layers may contain  significant amounts of He which,  at the effective
temperature    of   GD~362    would   be    almost   spectroscopically
invisible. Thus, GD~362 would still  be classified as a DA white dwarf
provided that some H is  present in its atmosphere.  Consequently, the
H/He ratio can be regarded  as a free parameter. However, the presence
of He in  a cool hydrogen-rich atmosphere affects  the surface gravity
determined  from   spectroscopy,  and  thus   the  mass  determination
\citep{Bergeron}. In  Fig. 1 we show three  almost identical synthetic
spectra representative  of GD~362 with various  assumed He abundances.
If He/H=10 is adopted then  $\log g=8.25$ is obtained ($M_{\rm WD}\sim
0.8 \, M_{\sun}$) whereas if  we adopt He/H=1 then the surface gravity
turns  out to  be $\log  g=8.72$. This  corresponds to  a mass  of the
primary of  $M_{\rm WD} \sim 1.0  \, M_{\sun}$, which  can be obtained
from  the coalescence  of  a  $0.4 +  0.8\,  M_{\sun}$ binary  system.
Additionally,  in this  case the  largest abundances  of  the relevant
elements are  obtained.  Thus, we choose  the $0.4 +  0.8 \, M_{\sun}$
simulation  as    our  reference   model, although  reasonable results
can be obtained adopting other masses.   In  passing,  we  note   that
nevertheless the  He abundance is rather uncertain  since equally good
fits  to the observed  spectrum of  GD~362 can  be obtained  with very
different  He abundances.   Thus, the  mass of  GD~362 is  also rather
uncertain.  More  importantly, if the mechanism  producing the unusual
photospheric abundance pattern of GD~362 were to be accretion from the
inner regions  of the disk  --- which are  C-- and O--rich  --- atomic
lines of  CI, and C$_2$ molecular  bands should be  rather apparent in
the  spectrum.  But the  strength of  these spectral  features depends
very  much  on  the  adopted  He abundance,  because  the  atmospheric
pressure  and the opacity  also depend  very much  on the  H/He ratio,
which is rather uncertain.

\begin{figure}
\begin{center}
\includegraphics[scale=.35]{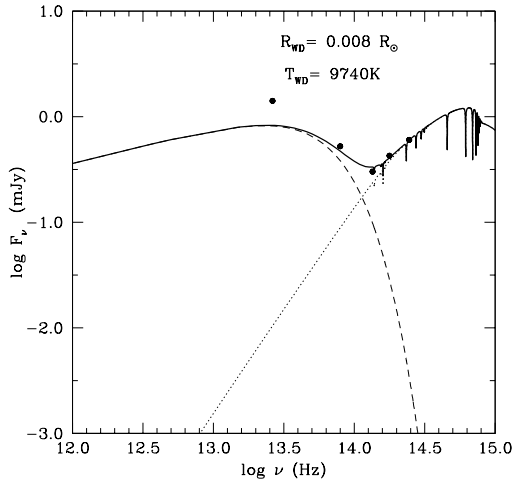}
\caption{Spectral energy distribution of  GD~362. The figure shows the
         spectral energy distribution of GD~362. The dotted line shows
         the spectrum of a  white dwarf with and effective temperature
         of 9740 K  and $\log g=8.72$, which corresponds  to a mass of
         about one solar mass, the dashed line shows the spectrum of a
         passive flat, opaque dust disk and the solid line depicts the
         composite spectrum. The observational data were obtained from
         \cite{Becklin}.}
\end{center}
\end{figure}

In  order to know  whether the  chemical abundances  of GD~362  can be
reproduced by direct  accretion from the keplerian disk  we proceed as
follows. Given  the surface gravity  and the effective  temperature of
our model we  compute the luminosity, the radius  and the cooling time
of  the  white   dwarf  according  to  a  set   of  cooling  sequences
\citep{Salaris}.   We obtain  respectively  $\log(L_{\rm WD}/L_{\sun})
\simeq -3.283$, $\log(R_{\rm WD} /R_{\sun})\simeq -2.096$, and $t_{\rm
cool}\simeq 2.2$~Gyr.   Hence, in this scenario GD~362  has had enough
time from  the moment in  which the merger  occurred to cool  down, to
accrete most of the C--  and O--rich region, settle down the accretion
disk, and to form dust.  Additionally, the central white dwarf has had
time enough to  accrete (at a rate much  smaller than the Bondi--Hoyle
accretion  rate) the small  amount of  hydrogen from  the ISM  to show
spectroscopic hydrogen features. We  further assume that the accretion
luminosity:

\begin{equation}
L_{\rm acc}=\frac{G\dot M M_{\rm WD}}{R_{\rm WD}}
\end{equation}

\noindent is, in  the worst of the cases,  smaller than the luminosity
of the white dwarf. This provides  us with an (extreme) upper limit to
the  accretion  rate, which  turns  out  to  be $1.3\times  10^{-13}\,
M_{\sun}\, {\rm yr}^{-1}$. Next, we assume that the abundance of Ca is
the  result  of the  equilibrium  between  the  accreted material  and
gravitational diffusion:

\begin{equation}
\dot M X_{\rm disk}=\frac{M_{\rm env}X_{\rm obs}}{\tau_{\rm diff}}
\end{equation}

\noindent where $X_{\rm disk}$ is the abundance in the accretion disk,
$X_{\rm obs}$ is the photospheric abundance, $M_{\rm env}$ is the mass
of  the envelope  of GD~362  and  $\tau_{\rm diff}$  is the  diffusion
timescale. The diffusion timescale of Ca for H--rich atmospheres is of
the order of a few  years. However, the accreted material is He--rich,
so  the diffusion  timescale is  probably more  typical of  a He--rich
envelope,  which is  much  larger \citep{Paquette},  of  the order  of
$\tau_{\rm diff}  \sim 1.5 \times  10^4$~yr.  Unfortunately, diffusion
timescales  for  mixed H/He  envelopes  do  not  exist.  However,  the
diffusion characteristic  times scale as  $\tau_{\rm diff}\propto \rho
T^{-1/2}g^{-2}$ \citep{Alcock}.  We  have computed detailed atmosphere
models  for  pure H,  He/H=1  and  He/H=10  and scaled  the  diffusion
timescale using the  values of the density and  the temperature at the
base of the convective zones and the appropriate chemical composition.
For our fiducial composition (He/H=10) we obtain $\tau_{\rm diff} \sim
8.5 \times 10^3$~yr.  From this we obtain the mass of the region where
diffusion  occurs, which turns  out to  be $M_{\rm  env}\sim 7.2\times
10^{-9}  \, M_{\sun}$,  which is  much smaller  than that  obtained by
accretion from  the interstellar medium at  the Bondi--Hoyle accretion
rate $(\sim  1.5 \times  10^{-6} M_{\sun}$).  Hence,  the photospheric
abundances of GD~362 can be successfully explained by direct accretion
from the surrounding disk.

Now we assess  whether the flux from the accretion  disk can be fitted
by the  results of our SPH  simulations. In order to  compute the flux
radiated away  from the  system two contributions  must be  taken into
account.  The  first one  is the expected  photospheric flux  from the
star, for which we use the spectral energy distribution ($B_{\rm WD}$)
of  a white  dwarf of  mass $1\,  M_{\sun}$, at  $T_{\rm  eff} \approx
9740$~K:

\begin{equation}
F_{\rm WD} = \pi \left(\frac{R_{\rm WD}}{D_{\rm WD}}\right)^2 
B_{\rm WD}(T_{\rm eff}),
\end{equation}

Given the luminosity of our model and the apparent magnitude of GD~362
we obtain a distance of $D_{\rm WD}=33$~pc. The second contribution to
the  total flux  comes from  the  emission of  the disk,  which for  a
passive flat, opaque dust disk is \citep{Chiang,Jura}:

\begin{eqnarray}
F_{\rm disk} \simeq&~& 12 \pi^{1/3} \cos i 
\left( \frac{R_{\rm WD}}{D_{\rm WD}}\right)^2
\left(\frac{2k_{\rm B}T_{\rm s}}{3h\nu}\right)^{8/3}\nonumber\\
&~&\left(\frac{h\nu^3}{c^2}\right)
\int^{x_{\rm out}}_{x_{\rm in}}\frac{x^{5/3}}{e^x-1}dx
\end{eqnarray}

\noindent where $i$ is the inclination  of the disk (which we adopt to
be  face--on),  $x_{\rm  in}=h\nu/k_{\rm  B}T_{\rm  in}$  and  $T_{\rm
in}=1200$~K  is the  condensation temperature  of silicate  dust.  The
outer  radius is taken  from the  results of  our SPH  simulations and
turns out  to be $R_{\rm  out} \approx 1  \, R_{\sun}$. The  result is
displayed in Fig. 2. The dots are the observational data for GD~362.

\begin{figure}
\begin{center}
\includegraphics[scale=.35]{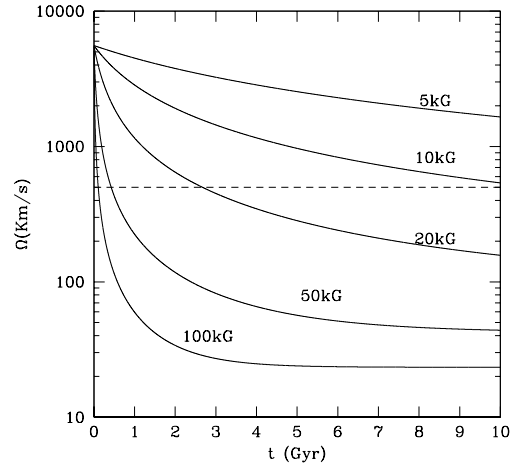}
\caption{Evolution  of  the  rotational  velocity  for  several  field
         strengths,  the  observational  upper  limit is  shown  as  a
         horizontal dashed line.}
\end{center}
\end{figure}

The proposed scenario has apparently two weak points. The first one is
that infrared observations indicate the presence of SiO. This requires
that O should be more abundant than C in order to form it. However our
simulations  show that  the  ratio of  C to  O  is a  function of  the
distance to the primary and, in  some regions of the disk the ratio is
smaller than  1, allowing  for the formation  of SiO in  the accretion
disk. Furthermore, after  2.2~Gyr of evolution the  resulting disk has
had time  to form  planets or asteroids  with the  subsequent chemical
differentiation.

The second  apparent drawback of the  model is that  the central white
dwarf rotates  very fast. However, an unobservable  magnetic field can
brake  down  the central  star  to  acceptable  velocities. Using  the
observed spectrum  of GD~362 it is  possible to set an  upper limit to
the  rotation speed  of $v\sin i \lesssim 500$~km~s$^{-1}$.  We assume
that  the central  white dwarf  has a  weak magnetic  field,  $B$. The
magnetic torques that lead to spin--down are caused by the interaction
between the white dwarf and the surrounding disk. The evolution of the
angular velocity due to the  coupling of the white dwarf magnetosphere
and the disk is given by \citep{Livio3,Armitage,Benacquista}:

\begin{eqnarray}
\dot{\Omega} &=&  -\frac{2\mu^2\Omega^3}{3I c^3}\sin^2{\alpha} + 
\frac{\mu^2}{3I}\left(\frac{1}{R_{\rm m}^3} -
\frac{2}{(R_{\rm c} R_{\rm m})^{3/2}}\right)\nonumber\\
&+&\frac{\dot{M}R^2_{\rm m}\Omega}{I}
\end{eqnarray}

\noindent where $\mu=BR^3_{\rm WD}$, $R_{\rm m}$ is the magnetospheric
radius of  the star,  $I$ is  the moment of  inertia, $\alpha$  is the
angle between  the rotation  and magnetic axes  (which we adopt  to be
$30^\circ$) and

\begin{equation}
R_{\rm c}=\left(\frac{GM_{\rm WD}}{\Omega^2}\right)^{1/3}
\end{equation}

\noindent is the corotation radius.  The first term in this expression
corresponds to  the magnetic dipole radiation emission,  the second to
the  disk--field coupling  and the  last one  to the  angular momentum
transferred from  the disk to  the white dwarf.  The  magnetic linkage
between the star and the disk leads to a spin--down torque on the star
if  the  magnetospheric  radius   is  large  enough  relative  to  the
corotation radius:

\begin{equation}
\left(\frac{R_{\rm m}}{R_{\rm c}}\right)\ge 2^{-2/3}
\end{equation}

We  adopt $R_{\rm  m}=R_{\rm  c}$.  Solving  numerically the  previous
differential equation  with the  appropriate parameters for  our case,
the evolution of the rotation velocity is shown in figure 3. As can be
seen a weak  magnetic field of about  50 kG is able to  brake down the
white dwarf to velocities below the observational limit. This magnetic
field is  much smaller than the  upper limit of about  0.7 MG obtained
from the spectrum of GD~362. Hence, our scenario also accounts for the
low   rotational  velocity   of  GD~362,   without   adopting  extreme
assumptions.

\section{Conclusions}

We have shown that  the anomalous photospheric chemical composition of
the DAZ white  dwarf GD~362 and of the  infrared excess of surrounding
disk can be  quite naturally explained assuming that  this white dwarf
is the result of the coalescence of a binary white dwarf system.  This
scenario  provides   a  natural  explanation  of   both  the  observed
photospheric abundances  of GD~362 and of its  infrared excess without
the need to invoke extreme assumptions, like the accretion of a planet
or an asteroid, since direct  accretion from the disk surrounding disk
provides a self--consistent way of polluting the envelope of the white
dwarf with the  required amounts of Ca, Mg, Si  and Fe. Moreover, this
last scenario can be also well accomodated within the framework of our
scenario given that the formation of planets and other minor bodies is
strongly  enhanced in  metal--rich  disks. Hence,  GD~362 could  be the
relic of a very rare event  in our Galaxy: the coalescence of a double
white dwarf binary system. 

\acknowledgments

This   work  has   been  partially   supported  by   the   MEC  grants
AYA05--08013--C03--01 and  02, by the  European Union FEDER  funds, by
the  AGAUR  and  by  the  Barcelona  Supercomputing  Center  (National
Supercomputer Center).   This work was  also supported in part  by the
NSERC (Canada).   P.  Bergeron is  a Cottrell Scholar of  the Research
Corporation.

\end{document}